\documentclass[sigconf,nonacm]{acmart}
\makeatletter
\def\@ACM@authorsperrow{5} 
\makeatother

\usepackage{multirow}
\usepackage[normalem]{ulem}
\usepackage{makecell}
\AtBeginDocument{%
  }
\renewcommand\footnotetextcopyrightpermission[1]{}

\author{Zining Wang}
\affiliation{%
  \institution{Northeastern University}
  \city{Boston}
  \country{USA}
}
\author{Jian Gao}
\affiliation{%
  \institution{Northeastern University}
  \city{Boston}
  \country{USA}
}
\author{Weimin Fu}
\affiliation{%
  \institution{Kansas State University}
  \city{Manhattan}
  \country{USA}
}
\author{Xiaolong Guo}
\affiliation{%
  \institution{Kansas State University}
  \city{Manhattan}
  \country{USA}
}
\author{Xuan Zhang}
\affiliation{%
  \institution{Northeastern University}
  \city{Boston}
  \country{USA}
}
\begin{document}

\title{AnalogSAGE: \uline{S}elf-evolving Analog Design Multi-\uline{A}gents with Stratified Memory and \uline{G}rounded \uline{E}xperience}

\begin{abstract}
Analog circuit design remains a knowledge- and experience-intensive process that relies heavily on human intuition for topology generation and device parameter tuning. Existing LLM-based approaches typically depend on prompt-driven netlist generation or predefined topology templates, limiting their ability to satisfy complex specification requirements. We propose \textbf{AnalogSAGE}, an open-source self-evolving multi-agent framework that coordinates three-stage agent explorations through four stratified memory layers, enabling iterative refinement with simulation-grounded feedback. To support reproducibility and generality, we release the source code. Our benchmark spans ten specification-driven operational amplifier design problems of varying difficulty, enabling quantitative and cross-task comparison under identical conditions. Evaluated under the open-source SKY130 PDK with \texttt{ngspice}, AnalogSAGE achieves a \textbf{10$\times$} overall pass rate, a \textbf{48$\times$} Pass@1, and a \textbf{4$\times$} reduction in parameter search space compared with existing frameworks, demonstrating that stratified memory and grounded reasoning substantially enhance the reliability and autonomy of analog design automation in practice.
\end{abstract}
\maketitle
\begin{figure}[]
	\centering
	\includegraphics[width=0.45\textwidth]{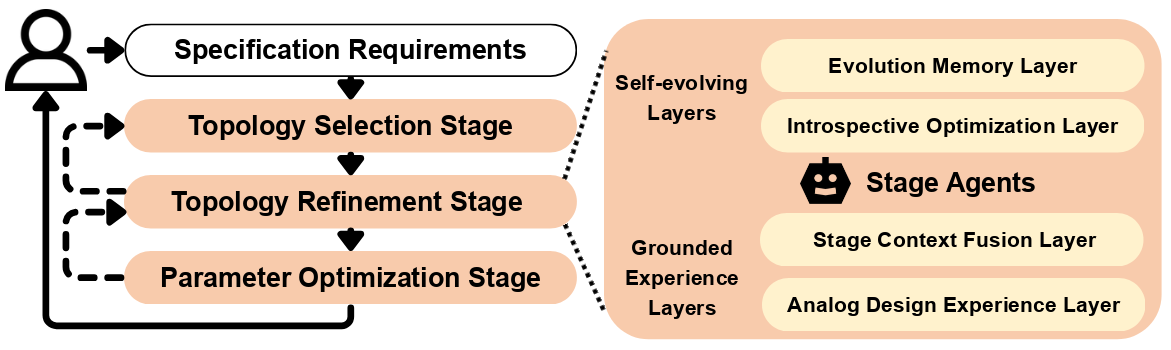}
	\caption{Overview of the proposed framework with three main stages: topology selection, topology refinement, and parameter optimization. The dashed line indicates closed-loop feedback. Each stage agent is supported by four memory layers that provide essential context for the LLM to perform iterative design improvements.}
	\label{fig:methodology_overview}
\end{figure}

\section{Introduction}

Analog circuit design is regarded as one of the most knowledge-intensive and experience-driven processes in electronic design automation (EDA)~\cite{guo2025toward}.
Unlike digital design, which benefits from structured synthesis and verification pipelines~\cite{AMURU2023102048}, analog design requires nuanced trade-offs among specification requirements that are traditionally guided by human intuition.
Despite decades of research in automation, fully autonomous analog synthesis remains elusive due to the enormous topology search space, nonlinear interactions among device parameters, and the absence of scalable reasoning frameworks capable of interpreting design intent~\cite{electronics11030435,9548117}.

Recent advances in large language models (LLMs) enable human-like reasoning in analog design automation as they can interpret textual specifications, retrieve knowledge, and propose design strategies. However, directly applying LLMs to analog circuit design still remains challenging~\cite{guo2025toward,akbari2025circuitsense}. 
Prompt-based approaches such as SPICEPilot~\cite{vungarala2024spicepilot} often generate structurally invalid circuits. For example, they may produce incorrect device connections. A key reason is that these methods lack feedback-driven refinement iterations. In contrast, template- and library-based methods such as Artisan~\cite{chen2024artisan} and AnalogXpert~\cite{zhang2025analogxpert} are generally more robust. However, they still struggle to generalize beyond predefined patterns, which limits their flexibility.
Some frameworks~\cite{lai2025analogcoder,chen2025menter} incorporate simulation results into the feedback loop to mitigate this issue, yet they rely on simplified device models, preventing accurate performance evaluation under practical PDK conditions. More fundamentally, existing methods lack persistent cross-task memory and simulation-grounded experience, leading to brittle long-horizon reasoning capabilities and reduced reliability under stringent analog specifications.

To address these challenges, this work presents AnalogSAGE, which enables LLM to evolve its design capability by generating new circuit topologies and learning from failed design tasks through simulation feedback~\cite{yang2025zerosim}, similar to human designers.
We design a stratified context mechanism to selectively preserve stage-relevant information while discarding irrelevant content, thereby maximizing effective context usage.
Numerical optimization methods are integrated with LLM-based reasoning to provide accurate sizing and bias parameters under grounded simulation. We release the implementation of the framework to facilitate fair comparison and accelerate the research on LLM-driven analog design.\footnote{Source code is available at \url{https://github.com/xz-group/AnalogSAGE}.}

\textbf{The main contributions of this work are as follows:}
\begin{itemize}
	\item \textbf{Self-evolving analog design multi-agent framework} that continuously enhances reasoning and design capability through accumulated knowledge, enabling transferable domain expertise and strategy refinement from both successful and failed design iterations.
	\item \textbf{A stratified memory architecture} that organizes and condenses essential design knowledge through four structured memory layers to support reasoning across different design stages within limited context windows.
	\item \textbf{A closed-loop framework with grounded experience} that unifies topology selection, refinement, and parameter optimization under simulator feedback. The framework is evaluated in a set of specification-driven design tasks using the SKY130 PDK, achieving a $10\times$ pass rate, a $48\times$ Pass@1, and a $4\times$ reduction in the parameter search space.
\end{itemize}

\section{Background}

\subsection{Analog design automation}

Traditional analog circuit automation has long depended on expert knowledge and manual optimization. Early design automation tools relied on predefined topology libraries selected by human designers, focusing mainly on parameter sizing rather than structural innovation. Subsequent AI-based methods, including reinforcement learning~\cite{zhao2020deep,9576505} and graph neural networks~\cite{liu2021parasitic,dong2023cktgnn}, expanded automation to limited topology generation and device tuning. Despite these advances, most works are limited to simplified device models, constrained search spaces, and expert-crafted heuristics. Automating topology synthesis remains challenging \cite{10937153} due to the vast design space, nonlinear performance trade-offs, and the nuanced judgment that human designers intuitively apply.

\subsection{Context management in AI agent}

Effective use of LLMs in complex design workflows requires careful management of limited context windows through isolation, compression, and reflection\cite{mei2025survey}. Isolated contexts allow agents to focus on specific subtasks without interference, compressed contexts summarize essential details to conserve space, and reflective contexts capture insights or mistakes from prior reasoning to improve future performance~\cite{chang2025sagallm}. Techniques such as summarization, pruning, and retrieval-based re-injection help maintain coherence across multi-step reasoning while preventing information overload~\cite{packer2023memgpt}. Through dynamic distillation and reuse of relevant information, the agent sustains continuity, recalls prior design choices, and applies analogical reasoning to new problems~\cite{lu2025scaling}.

Recent works such as ReAct~\cite{yao2022react} and Reflexion~\cite{shinn2023reflexion} demonstrate concrete implementations of reflective context mechanisms, combining action traces with self-evaluation to enhance long-horizon reasoning. Their iterative feedback processes highlight the importance of structured context updates for robust, multi-step decision making.

\subsection{Agentic frameworks in circuit design}

Recent research has explored combining LLMs with circuit simulators for analog design automation. However, many existing approaches still rely on prompt engineering~\cite{vungarala2024spicepilot} or rule-based scripts~\cite{zhang2025analogxpert} without true feedback loops or specification-driven reasoning. Their validation is often conducted using idealized device models~\cite{chen2025menter} rather than practical PDK environments, limiting real-world applicability. Furthermore, context length and knowledge retention are underutilized, preventing effective cross-task learning and simulation-grounded feedback~\cite{lai2025analogcoder,chen2024artisan}. 

In addition, most prior systems lack rigorous quantitative evaluation, do not test generality across multiple specification-driven tasks, and rarely release complete code or experimental pipelines, making independent reproduction and fair comparison difficult.

This work addresses these limitations through a self-evolving, simulation-grounded framework that integrates multi-context fusion, introspective feedback, and numerical optimization, while providing open-source code.
\begin{figure*}[t]
	\centering
	\includegraphics[width=0.9\textwidth]{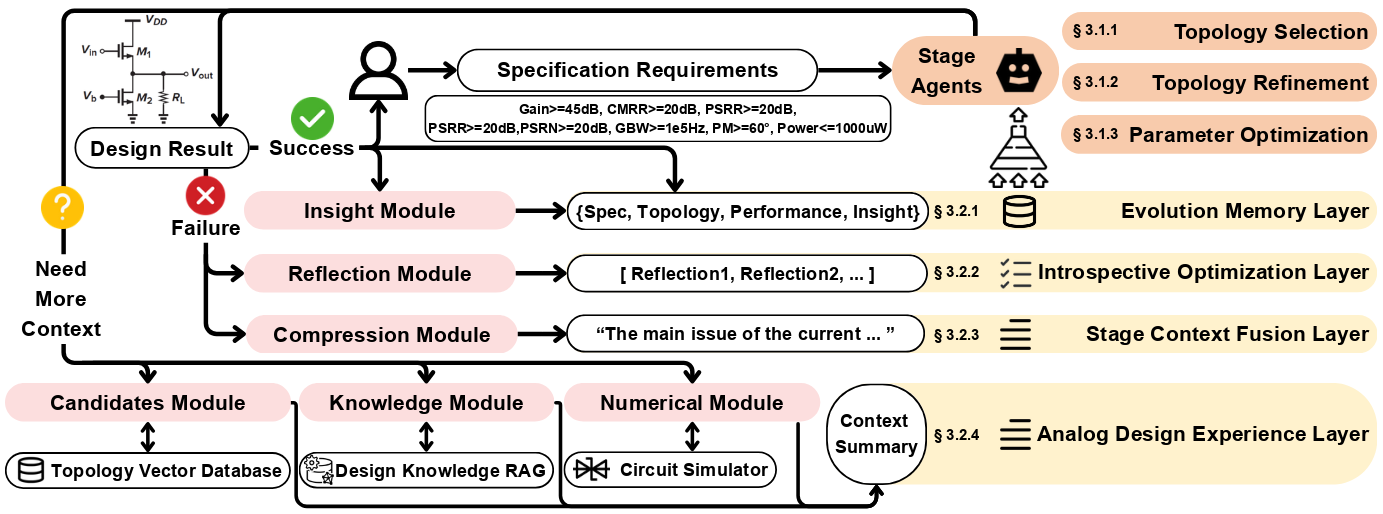}
	\caption{
	Interaction between the four context layers and the stage agents: Evolution Memory, Introspective Optimization, Stage Context Fusion, and Analog Design Experience. Each design iteration updates different parts of the memory layers based on the generated results.
	}
	\label{fig:context_overview}
\end{figure*}
\section{Methodology}

Our proposed framework, AnalogSAGE, translates analog circuit specifications into executable designs through an iterative, memory-augmented reasoning process. As illustrated in Figure~\ref{fig:methodology_overview}, this process is driven by three sequential agents: a \emph{Topology Selection Agent} proposes an initial architecture, a \emph{Topology Refinement Agent} modifies its structure to resolve design issues, and a \emph{Parameter Optimization Agent} determines the final numerical values for each device.

Crucially, the reasoning of the agents is supported by four \textit{Stratified Memory Layers}, which form the cognitive backbone of the system. \textit{Evolution Memory} transfers high-level insights across different design tasks, while \textit{Introspective Optimization} enables self-correction by learning from prior iterations within the same task. \textit{Stage Context Fusion} maintains a compressed summary of the current design state, and \textit{Analog Design Experience} grounds the agents with verified domain knowledge.

This tight integration of agents and memory allows \textit{AnalogSAGE} to operate as a self-evolving system, continually improving its design capabilities across both iterations and tasks. Here, a design task is defined by a set of specifications that is considered a pass only when all requirements are satisfied after simulation. Because initial solutions may fail, multiple design iterations are often required within a single task to refine the solution until the specifications are met, or the design still fails after reaching the maximum number of iterations.

\subsection{Stage agents}

The design process is divided into three reasoning stages, each controlled by a stage agent. Agents operate with different objectives: selecting a topology, refining its structure, and determining numerical parameter values. All agents share the same memory layer organization, but receive different tool feedback and contextual information. Information flows forward across stages, while compressed feedback is passed backward through the memory system, forming a closed loop that can iteratively solve the design task.

\subsubsection{Topology Selection Agent}

The \emph{Topology Selection Agent} initiates the design process by translating textual performance specifications into a concrete circuit architecture in netlist format. To avoid generating invalid topologies, the agent first queries the \emph{Stratified Memory Layers} for contextual signals derived from past design tasks and previous iterations. Conditioned on this context, it formulates a natural language description of the target architecture. This description is passed to the \emph{Candidates Module}, which retrieves the best-matching validated topology from the database. This retrieval-based initialization ensures that the design process begins from a structurally sound foundation.

\subsubsection{Topology Refinement Agent}

Once a promising topology is identified, the design flow enters an iterative refinement stage. Based on the gathered context, the agent performs localized structural modifications, such as adjusting bias networks or compensation paths. The refined topology, or the original one if no modification is required, is then passed to the next stage. All refinements are intentionally constrained to local edits to ensure that the resulting topology remains structurally valid.

\subsubsection{Parameter Optimization Agent}

Once the topology is fixed, the framework switches the reasoning mode from structural decisions to numerical prediction. The agent first infers feasible parameter ranges for all devices in the circuit, including transistor dimensions, bias conditions, and passive elements. These ranges define the search space and call the \emph{Numerical Module } to run Bayesian optimization and circuit simulation to evaluate candidate parameter sets. The resulting performance metrics are fed back into the \emph{Stratified Memory Layers}, allowing the agent to tighten the parameter bounds in subsequent iterations and progressively converge toward a specification-satisfying solution.

\subsection{Stratified memory layers}

The four memory layers provide the contextual backbone for the framework. They allow the stage agents to reason with both long-term transferable expertise and short-term task feedback. As shown in Figure ~\ref{fig:context_overview}, each layer contributes a different form of context: the \emph{Evolution Memory Layer} stores success design insights shared across tasks, the \emph{Introspective Optimization Layer} records failure reflections within a task, the \emph{Stage Context Fusion Layer} compresses intermediate reasoning traces, and the \emph{Analog Design Experience Layer} exposes tool-level and domain-specific knowledge outside the LLM. Working together, these layers enable iterative refinement without exceeding model context limits, and allow the system to improve over repeated deployments.

\subsubsection{Evolution Memory Layer}

It provides long-term transferable knowledge across design tasks. After each successful task, the \emph{Insight Module} generates a concise summary of the underlying design rationale, including effective topology choices, compensation strategies, and reasoning heuristics that led to convergence. Each summary is stored together with the associated specification, topology, and performance results.

When a new task arrives, the framework retrieves entries and injects them into the prompt as contextual guidance. Over time, this iterative accumulation and reuse of insights forms a transferable representation of analog design knowledge, enabling cross-task generalization and reducing the number of iterations required to meet new specifications, as it can learn from prior successful designs to have a higher chance of picking the right solution path.

\subsubsection{Introspective Optimization Layer}

While the \emph{Evolution Memory Layer} supports cross-task generalization, the \emph{Introspective Optimization Layer} enables within-task self-correction. After each unsuccessful design iteration, the \emph{Reflection Module} analyzes the reasoning trace and simulation results, and produces a concise reflection explaining the failure. Typical causes include inappropriate topology selection, insufficient biasing, or conflicting specification trade-offs. Each reflection also provides a concrete corrective action in future iterations and is stored in the \emph{Introspective Optimization Layer}.

In subsequent iterations under the same design task, where the specification remains unchanged. The framework retrieves these reflections and injects them into the prompt, preventing repeated failure patterns. This mirrors the workflow of expert analog designers, who refine intuition through targeted failure analysis rather than restarting from scratch. By continually incorporating these reflections, the system accumulates task-specific optimization heuristics, forming a closed loop that improves iteration efficiency and reduces redundant exploration during analog synthesis.

\subsubsection{Stage Context Fusion Layer}

As the number of iterations increases within a single design task, the accumulated textual context, including LLM design history, retrieved knowledge, and simulation feedback, grows rapidly in length. Such expansion increases computational overhead and degrades reasoning quality due to the LLM’s limited context capacity. To mitigate this problem, the \emph{Stage Context Fusion Layer} employs a \emph{Compression Module} that condenses long reasoning traces into concise summaries while preserving essential performance outcomes and decision rationales in the original context.

The fused context is then passed backward across stages, allowing later stage results to inform earlier stage decisions. This mechanism forms a feedback bridge that maintains continuity across iterations, reduces redundant decision exploration, and enables progressively more focused reasoning. By continuously distilling and re-injecting contextual information, the system establishes a closed-loop refinement process without exceeding model context limits.

\subsubsection{Analog Design Experience Layer}

This layer grounds the LLM’s reasoning in verifiable analog design knowledge rather than relying solely on language priors. Instead of generating circuits purely from the model itself, the stage agents could use APIs to communicate with external tools whenever structural evidence, analog domain knowledge, or numerical validation is required. These interactions are routed through three modules: the \emph{Candidates Module}, which retrieves structurally valid topologies; the \emph{Knowledge Module}, which provides literature-derived design guidance; and the \emph{Numerical Module}, which returns simulation-based performance feedback. By integrating these tool results directly into the reasoning loop, the system forms a hybrid symbolic–numeric workflow in which every major design decision is supported by grounded evidence rather than speculative generation.

\paragraph{Candidates Module}
This module works with the \emph{Topology Vector Database}, which stores analog circuit topologies paired with their netlists and concise two-sentence textual summaries. The first sentence describes the global architectural style (e.g., folded-cascode, two-stage Miller), while the second explains the internal stage structure and signal flow. These summaries are embedded into a vector database for semantic retrieval.

\begin{figure}[]
    \centering
    \includegraphics[width=0.4\textwidth]{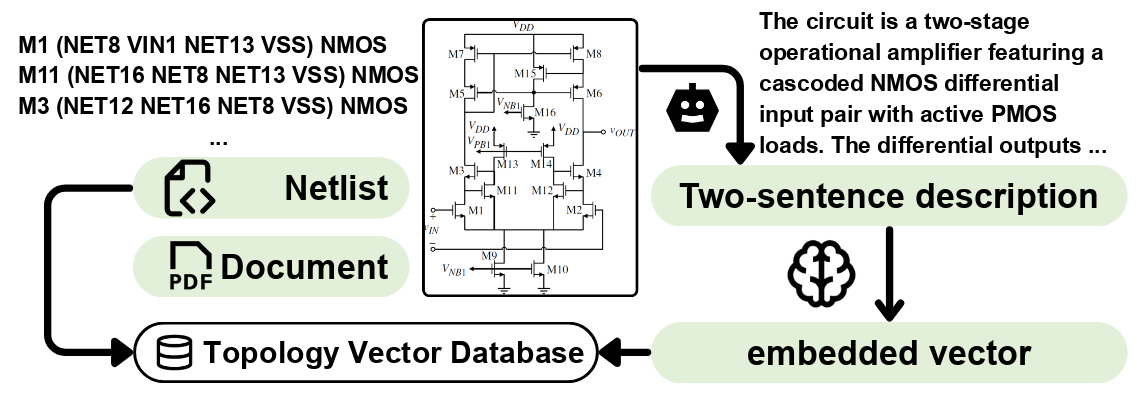}
    \caption{Construction workflow of the \emph{Topology Vector Database}. Each topology is converted into a two-sentence textual representation and stored as a key--value pair, where the embedded description serves as the key and the netlist as the value, enabling efficient retrieval.}
    \label{fig:topology}
\end{figure}

As shown in Figure~\ref{fig:topology}, each entry is constructed from two sources: the netlist and its original design document. The LLM parses the source text and generates the two-sentence description, which is then embedded into a vector representation. The database is organized as a key–value store, where the embedded description serves as the key and the netlist as the value, enabling efficient retrieval from textual queries. 

When triggered, the stage agent first produces a full natural-language description of the desired topology. The \emph{Candidates Module} then condenses this long description into the same two-sentence format used by the database and performs a cosine similarity search to obtain structurally relevant candidates. The retrieved results are further re-ranked according to constraints inferred from the agent’s original description, ensuring that the final ordering reflects the intended architectural direction rather than pure semantic similarity. The re-ranked topologies are returned to the design loop as valid structural hypotheses for subsequent refinement.

\paragraph{Knowledge Module}

This module operates on a Retrieval-Augmented Generation (RAG) system that enriches the LLM's reasoning with domain-specific knowledge extracted from human-authored analog design literature. All collected documents are segmented into semantically coherent sections, with non-informative content such as acknowledgements removed. The processed segments are then embedded to construct a \emph{Design Knowledge RAG}.

During the design process, whenever a stage agent requires theoretical or empirical guidance, such as specification trade-offs, the module is invoked. It automatically expands the original query into multiple sub-queries, searches the RAG, and retrieves the most relevant text segments. To prevent context overflow, the retrieved content is summarized into concise paragraphs tailored to the query, filtering out irrelevant information before being returned to the agent as auxiliary knowledge. This module ensures that the LLM can reason with up-to-date and contextually relevant analog knowledge without exceeding its context length limitations.

Together with the \emph{Candidates Module},  human-grounded analog experiences are inserted into the design reasoning process.

\paragraph{Numerical Module}

The \emph{Numerical Module} serves as the numerical optimization component of the design pipeline. It deploys Bayesian Optimization (BO) to optimize device parameters while accommodating diverse circuit topologies. Once triggered, the BO script iteratively samples parameter values, calls the \emph{Circuit Simulator}, evaluates performance metrics, and updates a surrogate model to identify the optimal configuration.

Unlike the LLM, whose reasoning over analog behavior may be approximate or speculative, circuit simulation provides ground-truth numerical feedback which therefore injects simulation-grounded performance signals into the design loop, ensuring that parameter decisions are guided by accurate device-level evidence rather than language-only predictions. 

\section{An Example Task Walk-through}

To demonstrate the workflow in AnalogSAGE, we use an operational amplifier (op-amp) design task with a set of spec requirements, as shown in Table~\ref{tab:spec_case_study}, which is intentionally designed to be challenging and exceeds the design space in the topology candidates database as an example to illustrate how AnalogSAGE generates a new circuit design.

\begin{figure}[]
	\centering
	\includegraphics[width=0.45\textwidth]{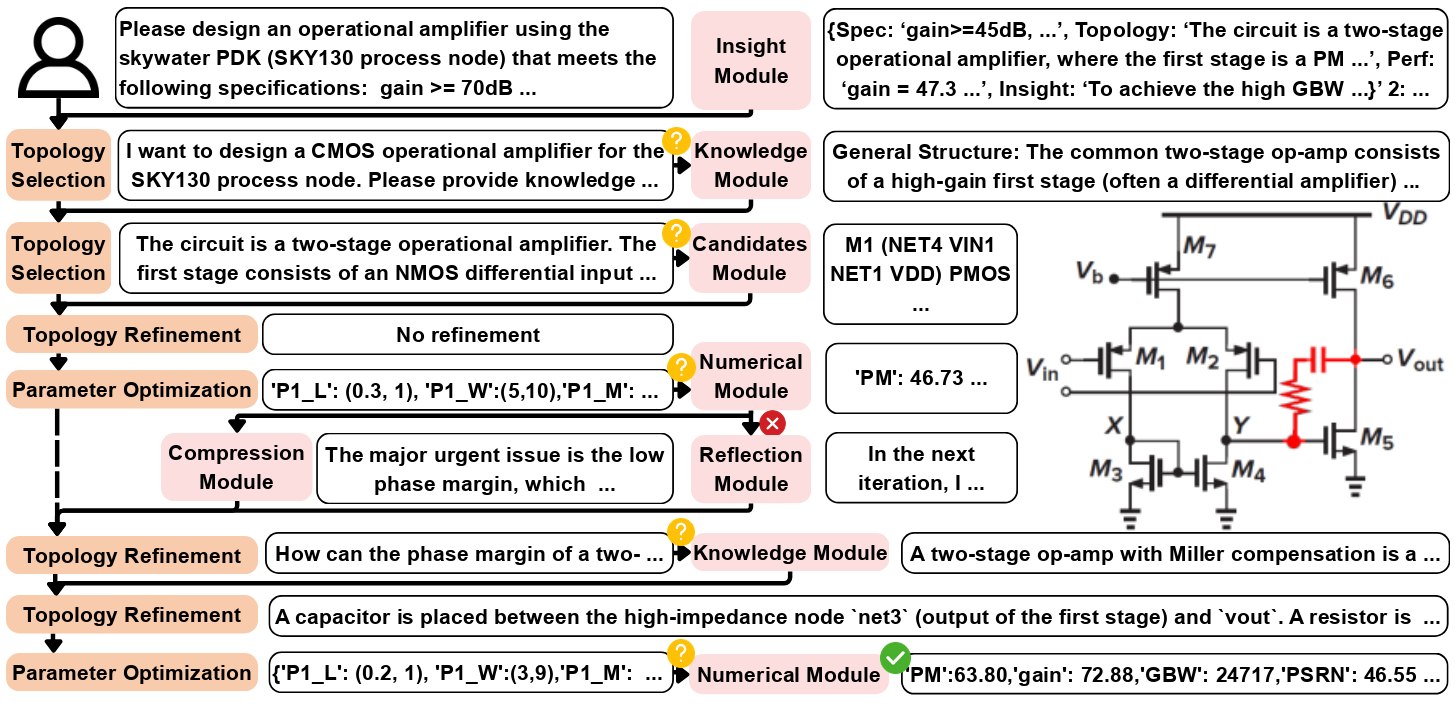}
	\caption{The log of the walk-through example. The refined structure is marked in red, adding Miller compensation with a nulling resistor connected to the output ($\text{V}_{\text{out}}$).}
	\label{fig:case_study_refinement}
\end{figure}

As shown in Figure~\ref{fig:case_study_refinement}, the \emph{Topology Selection Agent}, equipped with memory from the \emph{Insight Module} and augmented by the \emph{Knowledge Module}, identified an initial topology, which is a two-stage operational amplifier. This topology is then passed to the \emph{Topology Refinement Agent}, which performed no modification and directly forwarded the design to the \emph{Parameter Optimization Agent}, where parameter ranges were estimated and then provided to the \emph{Numerical Module} for numerical optimization. However, the resulting design failed to meet the target specifications, particularly the phase margin. As a failure case in the first iteration, the \emph{Compression Module} analyzed the cause of failure, which in this case was insufficient phase margin, and the \emph{Reflection Module} generated a reflection stating that phase margin must be improved. This feedback was returned to the \emph{Topology Refinement Agent}.

In the second iteration, the \emph{Topology Refinement Agent} received the failure analysis and reflection from the previous iteration and queried the \emph{Knowledge Module} again for improvement suggestions. The retrieved knowledge summarized that introducing Miller compensation would improve phase margin. Following this suggestion, the \emph{Topology Refinement Agent} performed refinement by adding a nulling resistor in series with the Miller capacitor. The modified topology was regenerated as a \texttt{PySpice} \cite{PySpice}  subcircuit and converted into an \texttt{ngspice} \cite{sourceforgeNgspice} compatible netlist, since the LLM is more familiar with Python-based circuit descriptions than raw SPICE syntax. 
Through iterative optimization in the \emph{Parameter Optimization Agent}, the system identified device parameters whose simulated performance satisfied all target specifications. This marked the design as successfully completed. The final result was then stored to support self-evolution and improve performance on future design tasks.

\begin{table}[]
	\centering
	\caption{Spec targets and simulation results in the example.}
	\label{tab:spec_case_study}
	\resizebox{\linewidth}{!}{
		\begin{tabular}{lccc c}
			\hline
			\textbf{Specifications} & \textbf{Unit} & \textbf{Design Target} & \textbf{ AnalogSAGE Result} \\
			\hline
			Gain                    & dB            & $\geq 70$              & 72.88                     \\
			GBW  & Hz            & $\geq 2 \times 10^5$   & 2.4 $\times 10^5$         \\
			Phase Margin            & deg           & $\geq 60$              & 63.80                     \\
			CMRR                    & dB            & $\geq 50$              & 54.57                     \\
			PSRR                    & dB            & $\geq 40$              & 43.88                     \\
			PSRN                    & dB            & $\geq 40$              & 46.55                     \\
			Power                   & $\mu$W        & $\leq 35$              & 30.35                     \\
			\hline
		\end{tabular}}
\end{table}

\section{Experiment Setup}

\subsection{Specification-driven design tasks}

To evaluate the proposed system, we selected op-amps as the target circuit family, since they represent a fundamental building block in analog design. As shown in Table~\ref{tab:spec_tasks}, we defined ten design tasks with gradually increasing difficulty, grouped into three levels.  Each task defines a combination of performance metrics, including power, DC gain (Gain), common-mode rejection ratio (CMRR), power supply rejection ratio (PSRR), gain-bandwidth product (GBW), phase margin (PM), and power supply noise rejection (PSRN). Tasks~1--5 correspond to relatively easy specifications under relaxed constraints, while Tasks~6--9 impose stricter limits on power, bandwidth, and noise rejection requirements. Task~10 is the most challenging case.

\begin{table}[]
	\centering
	\caption{Specification requirements for Tasks 1--10.}
	\label{tab:spec_tasks}
	\resizebox{\linewidth}{!}{
		\begin{tabular}{ccccccccc}
			\hline
\textbf{Level} &
\textbf{Task} &
\makecell{\textbf{Power}\\($\mu$W)} &
\makecell{\textbf{Gain}\\(dB)} &
\makecell{\textbf{CMRR}\\(dB)} &
\makecell{\textbf{PSRR}\\(dB)} &
\makecell{\textbf{GBW}\\(Hz)} &
\makecell{\textbf{PM}\\(deg)} &
\makecell{\textbf{PSRN}\\(dB)}
\\
			\midrule
			\multirow{5}{*}{Easy} 
			& 1  & $\leq 1000$  & $\geq 45$  & $\geq 20$  & $\geq 20$  & $\geq 1.0\times10^{5}$ & $\geq 60$  & $\geq 20$ \\
			& 2  & $\leq 1000$  & $\geq 45$  & $\geq 40$  & $\geq 40$  & $\geq 1.0\times10^{6}$ & $\geq 60$  & $\geq 40$ \\
			& 3  & $\leq 100$   & $\geq 80$  & $\geq 50$  & $\geq 50$  & $\geq 1.0\times10^{6}$ & $\geq 60$  & $\geq 50$ \\
			& 4  & $\leq 100$   & $\geq 80$  & $\geq 70$  & $\geq 70$  & $\geq 1.0\times10^{6}$ & $\geq 70$  & $\geq 70$ \\
			& 5  & $\leq 100$   & $\geq 100$ & $\geq 70$  & $\geq 70$  & $\geq 1.0\times10^{6}$ & $\geq 70$  & $\geq 70$ \\
			\hline
			\multirow{4}{*}{Medium} 
			& 6  & $\leq 10$    & $\geq 60$  & $\geq 70$  & $\geq 70$  & $\geq 5.0\times10^{5}$ & $\geq 50$  & $\geq 70$ \\
			& 7  & $\leq 10$    & $\geq 60$  & $\geq 70$  & $\geq 70$  & $\geq 5.0\times10^{5}$ & $\geq 50$  & $\geq 70$ \\
			& 8  & $\leq 1$     & $\geq 25$  & $\geq 55$  & $\geq 55$  & $\geq 1.0\times10^{5}$ & $\geq 45$  & $\geq 55$ \\
			& 9  & $\leq 2$     & $\geq 45$  & $\geq 30$  & $\geq 30$  & $\geq 1.0\times10^{6}$ & $\geq 45$  & $\geq 30$ \\
			\hline
			Hard & 10 & $\leq 5$  & $\geq 80$  & $\geq 30$  & $\geq 30$  & $\geq 1.0\times10^{6}$ & $\geq 45$  & $\geq 30$ \\
			\hline
		\end{tabular}}
\end{table}

\subsection{Baselines from prior works}

To ensure a fair and representative evaluation, we compared our method against two opensource frameworks and adapted one method to our framework.

\begin{itemize}
    \item \textbf{SPICEPilot} ~\cite{vungarala2024spicepilot} focuses on automatic SPICE netlist generation and verification through a PySpice-based benchmark. It integrates LLM-driven code synthesis, simulation, and rule-based correctness checks under zero-shot prompting.
    \item \textbf{AnalogCoder-Pro} ~\cite{lai2025analogcoder} employs one-shot prompting with multiple iterations using PySpice simulations to generate sized topology results.
    \item \textbf{Artisan-style SAGE} is based on the Artisan ~\cite{chen2024artisan}, which integrates several predefined design patterns and instructs the LLM to generate designs based on given candidates. Since their codebase and candidate sets are not publicly available, we construct our own candidate library and embed it into the prompts. We further apply the Numerical Module for parameter optimization.
\end{itemize}
\subsection{Evaluation metrics}

We employed three complementary metrics to assess design success, reasoning efficiency, and optimization effectiveness: \textit{Pass@k}, \textit{Average Iterations}, and \textit{Normalized Parameter Search Space}.

\paragraph{Pass@k.}
The \textit{Pass@k} metric measures the probability that at least one correct design appears among $k$ generated trials:$\text{Pass@k} = 1 - \frac{\binom{n-c}{k}}{\binom{n}{k}}$,
where $n$ is the total number of trials and $c$ is the number of correct ones.
In our experiments, we use $n = 5$ and $k = 1$.

\paragraph{Average Iterations.}
For iterative frameworks, we report the average number of reasoning iterations needed to reach a valid design. Fewer iterations indicate more efficient reasoning and stronger convergence.

\paragraph{Normalized Parameter Search Space.}
We compute the parameter search range selected by the agent and normalize it to the full search space (set to 1). Smaller values indicate tighter parameter constraints and more efficient optimization.

\subsection{Agent framework configuration}

We constructed the \emph{Topology Database} by labeling 50 representative topology candidates and built the \emph{Design Knowledge RAG} using approximately 10,000 analog design papers collected from ISSCC proceedings \cite{issccISSCCInternational} spanning 1955 to 2024. All experiments were conducted using the \texttt{gemini-2.5-flash}~\cite{comanici2025gemini} and \texttt{text-embedding-3-large}~\cite{neelakantan2022text} models via API access. Circuit simulations were performed using \texttt{ngspice} under the SKY130 PDK, adapted from the open-source AnalogGym script~\cite{10.1145/3676536.3697117}.
For fairness, we set the maximum number of reasoning iterations in AnalogSAGE to three, matching the iteration limit used in AnalogCoder-Pro.

\section{Experiment Results}

\begin{table}[]
    \centering
    \caption{Pass@1 comparison between proposed AnalogSAGE and baselines.}
    \label{tab:baseline_vs_full}
    \resizebox{\linewidth}{!}{
        \begin{tabular}{cccccccccccc}
            \hline
            \textbf{Framework} & \textbf{1} & \textbf{2} & \textbf{3} & \textbf{4} & \textbf{5} & \textbf{6} & \textbf{7} & \textbf{8} & \textbf{9} & \textbf{10} & \textbf{Avg.}\\
            \midrule
            \textbf{SPICEPilot \cite{vungarala2024spicepilot}}       & 0.0 & 0.0 & 0.0 & 0.0 & 0.0 & 0.0 & 0.0 & 0.0 & 0.0 & 0.0 & 0.0\\
            \textbf{AnalogCoder-Pro \cite{lai2025analogcoder}}  & 0.0 & 0.0 & 0.0 & 20.0 & 0.0 & 0.0 & 0.0 & 0.0 & 0.0 & 0.0 & 2.0 \\
            \hline
            \textbf{Artisan-style SAGE}          & \textbf{100.0} & 80.0 & 80.0 & 60.0 & \textbf{100.0} & 20.0 & 0.0 & 0.0 & 20.0 & 0.0 & 46.0 \\
            \textbf{AnalogSAGE} & \textbf{100.0} & \textbf{100.0} & \textbf{100.0} & \textbf{100.0} & \textbf{100.0} & \textbf{80.0} & \textbf{100.0} & \textbf{100.0} & \textbf{100.0} & \textbf{80.0} & \textbf{96.0}\\
            \hline
        \end{tabular}}
\end{table}

\begin{table}[]
    \centering
    \caption{Evaluation result under four configurations.}
    \label{tab:ablation_study_splithead}
    \resizebox{\linewidth}{!}{
    \begin{tabular}{cccccccccccc}
        \hline
        \textbf{Configuration} & \textbf{Metric} &
        \textbf{1} & \textbf{2} & \textbf{3} & \textbf{4} & \textbf{5} &
        \textbf{6} & \textbf{7} & \textbf{8} & \textbf{9} & \textbf{10} \\
        \hline

        \multirow{1}{*}{W/o IO and EM} & Pass@1
            & \textbf{100.0} & 80.0 & \textbf{100.0} & 40.0 & 80.0
            & 40.0 & 20.0 & 60.0 & 40.0 & 0.0 \\

        \multirow{1}{*}{W/o IO} & Pass@1
            & \textbf{100.0} & \textbf{100.0} & \textbf{100.0} & 80.0 & \textbf{100.0}
            & 20.0 & 40.0 & 40.0 & 60.0 & 0.0 \\

        \multirow{2}{*}{W/o EM} & Pass@1
            & \textbf{100.0} & \textbf{100.0} & \textbf{100.0} & \textbf{100.0} & \textbf{100.0}
            & 40.0 & 60.0 & 80.0 & \textbf{100.0} & 20.0 \\

        & Avg Iter
            & \textbf{1.0} & 1.2 & 1.4 & 1.4 & \textbf{1.2}
            & NA & NA & NA & 1.8 & NA \\

        \multirow{2}{*}{Full(AnalogSAGE)} & Pass@1
            & \textbf{100.0} & \textbf{100.0} & \textbf{100.0} & \textbf{100.0} & \textbf{100.0}
            & \textbf{80.0} & \textbf{100.0} & \textbf{100.0} & \textbf{100.0} & \textbf{80.0} \\

        & Avg Iter
            & \textbf{1.0} & \textbf{1.0} &\textbf{1.2} & \textbf{1.0} & \textbf{1.2}
            & NA & \textbf{1.2} & \textbf{1.4} & \textbf{1.0} & NA \\
        \hline
    \end{tabular}}
\end{table}

\subsection{Performance comparison}

Table~\ref{tab:baseline_vs_full} compares the proposed AnalogSAGE against baselines. SPICEPilot, which relies solely on zero-shot prompting without simulation feedback or iterative reasoning, fails on all tasks, yielding a 0\% pass rate and 0\% Pass@1. AnalogCoder-Pro introduces limited feedback but lacks grounded numerical optimization under the PDK. It occasionally discovers a correct design after many trials, achieving a 10\% pass rate and only a 2\% Pass@1. Artisann-style SAGE benefits from predefined circuit templates and simulation-guided parameter tuning, achieving a 70\% pass rate. However, its first-try success remains below 50\%, demonstrating weak generalization to strict specifications. In contrast, AnalogSAGE achieves a \textbf{100\% pass rate}, meaning every design task is eventually solved, and a \textbf{96.0\% Pass@1}, meaning 96\% of tasks succeed without requiring additional trials. This reflects a \textbf{10 $\times$} pass rate and a \textbf{48$\times$} in Pass@1 over the prior works, demonstrating stronger reliability, faster convergence, and better generalization across difficulty levels.

\subsection{Role of self-evolving memory layers}

To quantify the contribution of the proposed self-evolving memory design, we conducted experiments under four configurations.  
The lowest configuration disables both the \emph{Introspective Optimization Layer (IO)} and the \emph{Evolution Memory Layer (EM)}, denoted as \emph{W/o IO and EM}.  
The second removes only IO (\emph{W/o IO}), while the third removes only EM (\emph{W/o EM}).  
The full system keeps both layers enabled. All configurations were evaluated across ten design tasks, and the results are summarized in Table~\ref{tab:ablation_study_splithead}.

The system without either IO or EM can handle easy and medium tasks, but completely fails on the most challenging ones. It also exhibits unstable convergence, often requiring multiple trials to obtain a feasible solution.
Adding only EM improves performance on medium tasks by enabling cross-task knowledge transfer, which initializes the search closer to feasible operating regions. The gradual Pass@1 improvement (from 20\% to 60\% in Tasks~6--9) confirms that accumulated design priors benefit later tasks, although the system still fails on the hardest case due to the lack of adaptive correction.
In contrast, enabling only IO allows the system to eventually solve Task~10, demonstrating that introspective self-correction can repair failing design trajectories. However, its Pass@1 remains low (20\%), indicating that without insight priors the system lacks stability.

When both IO and EM are enabled, the framework achieves the best overall performance, reaching a 100\% pass rate and the highest Pass@1 with fewer iterations. This shows that the \emph{Evolution Memory Layer} provides informed initialization for new design tasks, while the \emph{Introspective Optimization Layer} adaptively corrects reasoning paths using grounded feedback. The two layers are complementary and jointly necessary for robust autonomous analog design.

\subsection{Parameter optimization efficiency}

Beyond reasoning accuracy, we also evaluated AnalogSAGE’s efficiency in parameter optimization. In this stage, the LLM predicts parameter ranges for each transistor and passive component. These predicted ranges directly shape the search space for the subsequent numerical optimizer. Tighter and more accurate ranges reduce the search space, accelerate convergence, and increase the likelihood of success.
Table~\ref{tab:sizing_LLM} shows the normalized parameter search space across different configurations where the smaller is better.  
Our AnalogSAGE achieves an average space of 0.26, indicating a \textbf{4$\times$ reduction} in parameter search space, which outperforms other three configurations.  
These results confirm that the \emph{Knowledge Module} provides physically consistent parameter bounds, while the \emph{Introspective Optimization Layer} further tightens feasible regions through reflective, simulation-driven feedback.
\begin{table}[]
    \centering
    \caption{Comparison of normalized parameter search space across different configurations.}
    \label{tab:sizing_LLM}
\resizebox{\linewidth}{!}{
\begin{tabular}{cccccccccccc}
    \hline
    \textbf{Method} &
    \textbf{1} & \textbf{2} & \textbf{3} & \textbf{4} & \textbf{5} &
    \textbf{6} & \textbf{7} & \textbf{8} & \textbf{9} & \textbf{10} & \textbf{Avg.} \\
    \hline

    W/o IO and Knowledge
        & 0.11 & 0.36 & 0.67 & 1.00 & 1.00
        & 1.00 & 0.62 & 0.31 & 0.37 & 0.29 & 0.57  \\

    W/o IO
        & 0.29 & 0.03 & 0.89 & \textbf{0.31} & \textbf{0.37}
        & 0.35 & 0.36 & 0.58 & 0.06 & \textbf{0.0008} & 0.32 \\

    W/o Knowledge
        & \textbf{0.03} & 0.37 & 0.64 & 1.00 & 1.00
        & 1.00 & 1.00 & \textbf{0.30} & 0.37 & 0.33 & 0.60 \\
    Full (AnalogSAGE)
        & 0.07 & \textbf{0.02} & \textbf{0.27} & 1.00 & 0.45
        & \textbf{0.32} & \textbf{0.05} & 0.37 & \textbf{0.04} & 0.003 & \textbf{0.26} \\

    \hline
\end{tabular}}
\end{table}

\section{Conclusion}

This work presents \textbf{AnalogSAGE} that integrates LLM-based self-evolving reasoning, stratified memory, and grounded experience. By coordinating three stage agents under a four-layer memory architecture, it tightly couples symbolic reasoning with numerical optimization in a closed-loop workflow. The results suggest that self-evolving, memory-structured agentic frameworks can capture and generalize analog design knowledge in ways that parallel human learning.
Across ten specification-driven op-amp design tasks, AnalogSAGE achieves a \textbf{10$\times$ pass rate}, a \textbf{ 48$\times$ pass@1}, and a \textbf{4$\times$ reduction in parameter search space}.
\bibliographystyle{ACM-Reference-Format}
\bibliography{references}
\end{document}